\documentclass[amsmath,pra,twocolumn,showpacs]{revtex4}
\usepackage{graphics}

\begin{document}     

\title{Optical excitation spectrum of an atom in a surface-induced potential}

\author{ Fam Le Kien,$^{1,*}$ S. Dutta Gupta,$^{1,2}$ and K. Hakuta$^{1}$} 
\affiliation{
$^1$Department of Applied Physics and Chemistry, 
University of Electro-Communications, Chofu, Tokyo 182-8585, Japan\\
$^2$School of Physics, University of Hyderabad, Hyderabad, India}

\date{\today}

\begin{abstract}
We study the optical excitation spectrum of an atom in the vicinity of a dielectric surface.
We calculate the rates of the total scattering and the scattering into the evanescent modes. With a proper assessment of the limitations, we demonstrate the portability of the flat-surface results to an experimental situation  with a nanofiber. The effect of the surface-induced potential on the excitation spectrum
for free-to-bound transitions is shown to be weak. On the contrary, the effect for  bound-to-bound transitions is significant leading to a large excitation linewidth, a substantial negative shift of the peak position, and a strong long tail on the negative side and a small short tail  on the positive side 
of the field--atom frequency detuning. 
\end{abstract}

\maketitle

The study of individual neutral atoms in the vicinity of dielectric and metal surfaces has gained renewed interest due to progress in atom optics and nanotechnology \cite{Lima,Oria2006,Kali,boundspon,our traps,cesium decay,absorption}. The ability of manipulating atoms near surfaces has established them as a tool for cutting-edge applications such as quantum computation \cite{Schlosser,Kuhr,Sackett}, atom chips \cite{Folman,Eriksson}, and probes, very sensitive to the surface-induced perturbations \cite{surface probe}. 
Recently, translational levels of an atom in a surface-induced potential have been 
studied \cite{Lima,Oria2006,Kali,boundspon}.
An optical technique for loading atoms into
quantum adsorption states of a dielectric surface has been suggested \cite{Lima,Oria2006}.
An experimental observation of the excitation spectrum of cesium atoms in quantum adsorption states of a nanofiber surface has been reported \cite{Kali}.
Spontaneous radiative decay of translational levels of an atom near a dielectric surface
has been investigated \cite{boundspon}. In this paper, we
study the optical excitation spectrum of an atom in a surface-induced potential.

We assume the whole space to be divided into two regions, namely,
the half-space $x<0$, occupied by a nondispersive nonabsorbing dielectric medium (medium 1), and
the half-space $x>0$,  occupied by vacuum (medium 2).
We examine an atom, with an upper internal level $e$ and a lower internal level $g$, moving in the empty half-space $x>0$. 

The potential energy of the surface--atom interaction is
a combination of a long-range van der Waals attraction $-C_3/x^3$ and
a short-range repulsion \cite{Hoinkes}. 
Here $C_3$ is the van der Waals coefficient.
We approximate the short-range repulsion by an exponential function $Ae^{-\alpha x}$,
where the parameters $A$ and $\alpha$ determine the height and range, respectively, of the repulsion.
The combined potential depends on the internal state of the atom (see Fig. \ref{fig1}), and
is presented in the form
$V_j(x)=A_j e^{-\alpha_j x}-C_{3j}/x^3$,
where $j=e$ or $g$ labels the internal state of the atom.
The potential parameters $C_{3j}$, $A_j$, and $\alpha_j$ depend 
on the dielectric and the atom. 
In numerical calculations, we use the parameters of fused silica, for the dielectric,
and the parameters of atomic cesium with the $D_2$ line, for the two-level atomic model.
According to Ref. \cite{boundspon}, 
the parameters of the ground- and excited-state potentials for the silica--cesium interaction are  
$C_{3g}=1.56$ kHz $\mu$m$^3$, $C_{3e}=3.09$ kHz $\mu$m$^3$, 
$A_g=1.6\times 10^{18}$ Hz, $A_e=3.17\times 10^{18}$ Hz, 
and $\alpha_g=\alpha_e=53$ nm$^{-1}$.

We introduce the notation $\varphi_{a}\equiv \varphi_{\nu_e}$ 
and $\varphi_{b}\equiv \varphi_{\nu_g}$ 
for the  eigenfunctions of the center-of-mass motion of the atom in the potentials
$V_e$ and $V_g$, respectively. 
They are determined by the stationary Schr\"{o}dinger equations
\begin{equation}
\left[-\frac{\hbar^2}{2m}\frac{d^2}{dx^2}+V_j(x)\right]\varphi_{\nu_j}(x)
=\mathcal{E}_{\nu_j}\varphi_{\nu_j}(x).
\label{e1}
\end{equation}
Here $m$ is the mass of the atom.
The eigenvalues $\mathcal{E}_{a}\equiv \mathcal{E}_{\nu_e}$ and 
$\mathcal{E}_{b}\equiv \mathcal{E}_{\nu_g}$ are the total center-of-mass energies
of the translational levels of the excited and ground states, respectively. 
These eigenvalues are the shifts of the energies of the translational 
levels from the energies of the corresponding internal states.
Without the loss of generality, we assume that
the center-of-mass eigenfunctions $\varphi_{a}$ and $\varphi_{b}$ are real functions.

%%%%%%%%%%%%%%%%%%%%%%% Figure 1
\begin{figure}[tbh]
\begin{center}
 \end{center}
\caption{Energies and wave functions of the center-of-mass motion of the ground- and excited-state atoms in the surface-induced potentials. 
The parameters of the potentials are 
$C_{3g}=1.56$ kHz $\mu$m$^3$, $C_{3e}=3.09$ kHz $\mu$m$^3$, 
$A_g=1.6\times 10^{18}$ Hz, $A_e=3.17\times 10^{18}$ Hz, 
and $\alpha_g=\alpha_e=53$ nm$^{-1}$.
The mass of atomic cesium $m=132.9$ a.u. is used.
We plot, for the excited state, two bound levels ($\nu=400$ and 415)
and, for the ground state, two bound levels ($\nu=281$ and 285) and  
one free level ($\mathcal{E}_f=4.25$ MHz).  
}
\label{fig1}
\end{figure}

We introduce the combined eigenstates $|a\rangle=|e\rangle\otimes|\varphi_{a}\rangle$
and $|b\rangle=|g\rangle\otimes|\varphi_{b}\rangle$, which are formed from
the internal and translational eigenstates.
The corresponding energies are 
$\hbar\omega_{a}=\hbar\omega_e+\mathcal{E}_{a}$
and $\hbar\omega_{b}=\hbar\omega_g+\mathcal{E}_{b}$. 
Here, $\omega_j$ with $j=e$ or $g$ is the frequency of the internal level $j$. 
Then, the Hamiltonian of the atom moving in the surface-induced potential 
can be represented in the diagonal form
$H_A=\sum_{a}\hbar\omega_{a}|a\rangle\langle a|+
\sum_{b}\hbar\omega_{b}|b\rangle\langle b|$.
We emphasize that the summations over $a$ and $b$ include both the discrete 
($\mathcal{E}_{a,b}<0$) and continuous ($\mathcal{E}_{a,b}>0$) spectra. 
The levels $a$ with $\mathcal{E}_{a}<0$ and the levels $b$ with $\mathcal{E}_{b}<0$
are called the bound (or vibrational) levels of the excited and ground states, respectively. 
In such a state, the atom is bound to the surface. 
It is vibrating, or more exactly, moving back and forth
between the walls formed by the van der Waals potential and the repulsion potential.
The levels $a$ with $\mathcal{E}_{a}>0$ and the levels $b$ with $\mathcal{E}_{b}>0$
are called the free (or continuum) levels of the excited and ground states, respectively.
The center-of-mass wave functions of the bound levels are normalized to unity.
The center-of-mass wave functions of the free levels are normalized to the delta function of energy.
For bound states, the center-of-mass wave functions $\varphi_{a}$ and $\varphi_{b}$ 
can be labeled by the quantum numbers $\nu_a$ and $\nu_b$, respectively. 
For free levels, the conventional sums over $a$ and $b$ must be replaced by the integrals
over $\mathcal{E}_{a}$ and $\mathcal{E}_{b}$, respectively.

Suppose that the atom is driven by a coherent plane-wave laser field 
$\mathbf{E}_1=[\mathcal{E} e^{-i(kx+\omega t)}\boldsymbol{\epsilon}
+\mathrm{c.c.}]/2$, propagating perpendicularly to the surface of the dielectric.
Here $\mathcal{E}$, $\boldsymbol{\epsilon}$, 
$\omega$, and $k=\omega/c$ are the envelope, the polarization vector, the frequency, and the wave number, respectively, of the laser field.
The field reflected from the surface is $\mathbf{E}_2=
-R[\mathcal{E} e^{i(kx-\omega t)}\boldsymbol{\epsilon}
+\mathrm{c.c.}]/2$, where $R=(n_1-1)/(n_1+1)$ is the reflection coefficient.
According to Ref. \cite{boundspon}, the time evolution of the density matrix $\rho$ 
of the atom is governed by the equations
\begin{eqnarray}
\dot{\rho}_{aa'}&=&\frac{i}{2}\sum_{l,b}(\Omega_{lab}\rho_{a'b}^* e^{-i\delta_{ab}t}
-\Omega_{la'b}^*\rho_{ab}e^{i\delta_{a'b}t})
\nonumber\\&&\mbox{} 
-\frac{1}{2}(\gamma_{a}+\gamma_{a'})\rho_{aa'},
\nonumber\\
\dot{\rho}_{ab}&=&
\frac{i}{2}\sum_{l,b'}\Omega_{lab'}\rho_{b'b}e^{-i\delta_{ab'}t}
-\frac{i}{2}\sum_{l,a'}\Omega_{la'b}\rho_{aa'}e^{-i\delta_{a'b}t}
\nonumber\\&&\mbox{} 
-\frac{1}{2}\gamma_{a}\rho_{ab},
\nonumber\\
\dot{\rho}_{bb'}&=&-\frac{i}{2}\sum_{l,a}(\Omega_{lab'}\rho_{ab}^*e^{-i\delta_{ab'}t}
-\Omega_{lab}^*\rho_{ab'}e^{i\delta_{ab}t})
\nonumber\\&&\mbox{}
+\frac{1}{2}\sum_{aa'}(\gamma_{aa'bb'}+\gamma_{a'ab'b})
e^{i(\omega_{bb'}-\omega_{aa'})t}{\rho}_{aa'},
\label{e2}
\end{eqnarray}
where the parameters $\gamma_{aa'bb'}$ and $\gamma_{a}\equiv\sum_b \gamma_{aabb}$ are the radiative decay coefficients \cite{boundspon}, the index  $l=1$ or 2 labels the incident or reflected field, respectively, 
the notation $\delta_{ab}=\omega-\omega_a+\omega_b$ stands for the detuning of 
the field frequency $\omega$ from the atomic transition frequency 
$\omega_{ab}=\omega_a-\omega_b$, 
and the quantities
\begin{eqnarray}
\Omega_{1ab}&=&\Omega F_{ab}(-k)=\Omega F_{ab}^*(k),\nonumber\\
\Omega_{2ab}&=&-R\Omega F_{ab}(k)
\label{e3}
\end{eqnarray}
are the Rabi frequencies of the fields $\mathbf{E}_{l=1,2}$ with respect to the translational-state transition
$a\leftrightarrow b$. 
Here, $\Omega=\mathcal{E}d_{eg}/\hbar$  is the Rabi frequency of the driving field 
with respect to the transition between the internal states $|e\rangle$ and $|g\rangle$, with 
$d_{eg}=\boldsymbol{\epsilon}\cdot\langle e|\mathbf{d}|g\rangle$ being the projection of the atomic dipole moment onto the polarization vector $\boldsymbol{\epsilon}$.
The coefficient
\begin{equation}
F_{ab}(k)=\langle \varphi_{a}|e^{ikx}| \varphi_{b}\rangle
\label{e4}
\end{equation}
is the overlapping matrix element that characterizes the transition between the center-of-mass states 
$|\varphi_{a}\rangle$ and $|\varphi_{b}\rangle$ with a transferred momentum of $\hbar k$.
Since $\varphi_{a}$ and $\varphi_{b}$ are real functions, we have $F_{ab}(-k)=F_{ab}^*(k)$.
In deriving Eqs. (\ref{e3}), we have neglected the cross decay coefficients  
$\gamma_{aa'}\equiv\sum_b\gamma_{aa'bb}$ with $a\not=a'$. Such coefficients are
small compared to the radiative linewidths $\gamma_a$ and $\gamma_{a'}$ when the refractive index $n_1$ of the dielectric is not large \cite{boundspon}.

\begin{figure}[tbh]
\begin{center} 
  \includegraphics{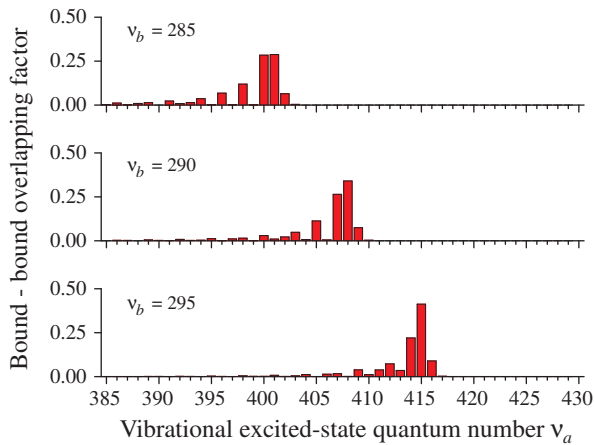}
 \end{center}
\caption{Frank--Condon factors $|F_{ab}(k)|^2$ for three bound ground-state levels $b$, with
$\nu_b=285$, 290, and 295, and various bound excited-state levels $a$, with $\nu_a$ in the range from
385 to 430. The light wavelength is $\lambda=852$ nm.
Other parameters  are as in Fig.~\ref{fig1}. 
}
\label{fig2}
\end{figure}

We plot in Fig. \ref{fig2} the Frank--Condon factors $|F_{ab}(k)|^2$ for three bound ground-state levels $\nu_b$ and various bound excited-state levels $\nu_a$. As seen from the figure, a deep bound ground-state level can be substantially coupled to a deep excited-state level.  
A deeper lower level can substantially overlap with a deeper upper level. 
The level $\nu_b=285$, with translational energy shift $\mathcal{E}_b=-54.39$ MHz,  can be substantially coupled
to the level $\nu_a=400$, with translational energy shift $\mathcal{E}_a=-132.84$ MHz. The difference $\mathcal{E}_a-\mathcal{E}_b=-78.45$ MHz is negative and large
compared to the natural linewidth $\gamma_0=5.25$ MHz of the $D_2$ line of atomic cesium.
In principle, both types of shifts, namely red (negative) shifts ($\mathcal{E}_a-\mathcal{E}_b<0$) 
and blue (positive) shifts ($\mathcal{E}_a-\mathcal{E}_b>0$), of the transition frequencies may be obtained.
Note that the van der Waals potential for the excited state is stronger than that for the ground state. Therefore, 
for a bound-to-bound transition with a substantial Frank--Condon factor, the energy shift of the excited-state level is usually larger than that of the ground-state level. 
Consequently, the  strong bound-to-bound transitions usually have red shifts.
Since the depths of the surface-induced potentials $V_g$ and $V_e$ are large (159.6 THz for $V_g$ 
and 316 THz for $V_e$), the range of the frequency shifts of the strong bound-to-bound transitions is broad,
and the shifts can reach large negative values. 

%%%%%%%%%%%%%%%%%%%%%%% Figure 2

\begin{figure}[tbh]
\begin{center}
  \includegraphics{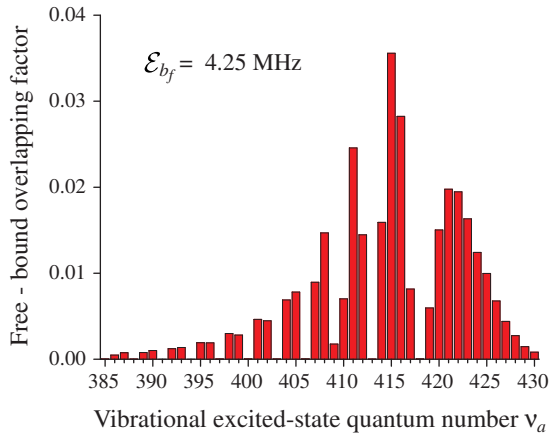}
 \end{center}
\caption{Frank--Condon factors $|F_{ab_f}(k)|^2$ for a free ground-state level $b_f$, 
with $\mathcal{E}_{b_f}=4.25$ MHz,
and various bound excited-state levels $a$, with $\nu_a$ in the range from 385 to 430. 
The parameters used for calculations are as in Fig.~\ref{fig2}. 
}
\label{fig3}
\end{figure}

We plot in Fig. \ref{fig3} the Frank--Condon factors $|F_{ab_f}(k)|^2$ for a free ground-state level $b_f$ and various bound excited-state levels $a$. The figure shows that
the values of the Frank--Condon factors $|F_{ab_f}(k)|^2$ are more substantial for shallow bound excited-state levels than for deep ones. Among the bound excited-state levels, the
level that is most strongly coupled to the free ground-state level $b_f$ is the level with 
the quantum number $\nu_a=415$. The translational energy shift of this level is $\mathcal{E}_a=-6.56$ MHz. This bound level of the excited state is rather shallow. Due to this fact, the frequency shift $\mathcal{E}_a-\mathcal{E}_{b_f}=-10.81$ MHz of the strongest free-to-bound transition from the free level $\mathcal{E}_{b_f}=4.25$ MHz 
 is negative but not large. We note that, when the temperature of the atomic system is low, the typical values of the free-level energy $\mathcal{E}_{b_f}$ are small. In this case, the range of the frequency shifts of substantial free-to-bound transitions is small compared to the range of the frequency shifts of substantial bound-to-bound transitions. 
 
%%%%%%%%%%%%%%%%%%%%%%% Figure 3

We now calculate the excitation spectrum of the atom in the surface-induced potential.
Assume that the field is so weak that the Rabi frequency 
$\Omega$ is much smaller
than the natural linewidth $\gamma_0$. We also assume that 
the atom is initially in a coherent mixtures of the ground-state levels $b$ with the weight
factors $\rho_{bb}=\rho_{bb}(0)$, which are the initial populations of the levels. We consider the adiabatic regime. We find from Eqs. (\ref{e2}) that the  population $\rho_{aa}$ of the excited-state level $a$ is given by
\begin{equation}
\rho_{aa}=\frac{s}{2}
\sum_{b}\frac{\gamma_0^2}{\gamma_a^2+4\delta_{ab}^2}|F_{ab}^*(k)-RF_{ab}(k)|^2\rho_{bb},
\label{e5}
\end{equation}
where $s=2|\Omega|^2/\gamma_0^2$ is the saturation parameter.
The rate of total scattering of light from the atom is given by
$\Gamma_{\mathrm{sc}}=\sum_a\rho_{aa}\gamma_a$. With the help of Eq. (\ref{e5}), we find
\begin{equation}
\Gamma_{\mathrm{sc}}=\frac{s}{2}
\sum_{ab}\frac{\gamma_0^2}{\gamma_a^2+4\delta_{ab}^2}\gamma_a|F_{ab}^*(k)-RF_{ab}(k)|^2\rho_{bb}.
\label{e6}
\end{equation}
We note that, due to the presence of the interface, the field in a mode can be either  a propagating light wave  or a bound evanescent wave \cite{Girlanda,Born}. 
An evanescent wave appears on the vacuum side of the interface 
when a light beam passing from the dielectric to the vacuum is totally internally reflected. 
Therefore, we can decompose 
the spontaneous emission rate as $\gamma_a=\gamma_a^{(\mathrm{ev})}
+\gamma_a^{(\mathrm{rad})}$, where $\gamma_a^{(\mathrm{ev})}$ and $\gamma_a^{(\mathrm{rad})}$
are the rates of spontaneous emission from the level $a$ into the bound evanescent modes and the propagating radiation modes, respectively \cite{boundspon}. Hence, we have 
$\Gamma_{\mathrm{sc}}=\Gamma_{\mathrm{sc}}^{(\mathrm{ev})}
+\Gamma_{\mathrm{sc}}^{(\mathrm{rad})}$, where
\begin{equation}
\Gamma_{\mathrm{sc}}^{(\mathrm{ev})}=\frac{s}{2}
\sum_{ab}\frac{\gamma_0^2}{\gamma_a^2+4\delta_{ab}^2}\gamma_a^{(\mathrm{ev})}|F_{ab}^*(k)-R F_{ab}(k)|^2\rho_{bb}.
\label{e7}
\end{equation}
is  the rate of scattering from the atom into the evanescent modes and
$\Gamma_{\mathrm{sc}}^{(\mathrm{rad})}=\Gamma_{\mathrm{sc}}-\Gamma_{\mathrm{sc}}^{(\mathrm{ev})}$ is the rate of scattering from the atom into the propagating radiation modes. 

All the above results were derived for a two-level atom moving in a potential induced by an infinite flat surface, and Eqs. (\ref{e6}) and (\ref{e7}) 
represent the rigorous expressions for the rates of 
the total scattering and the scattering into the evanescent modes (in the framework of the adiabatic approximation and the perturbation theory). However, with a proper assessment of the limitations, the above theory can be applied to typical experimental situations. As an example, we pick  
a recent experiment \cite{Kali}, whereby, the number of photons scattered from cesium atoms into the guided modes of a nanofiber was measured as a function of the frequency of the probe field. 
To apply our flat-surface results to the situation of a nanofiber, we use
several assumptions and approximations. 
First of all, we note that, in the case of an atom near a fiber, the effective potential for the center-of-mass motion of the atom along the radial direction
must include the centrifugal potential $U_{\mathrm{cf}}={\hbar^2(l_z^2-1/4)}/{2m r^2}$.
Here $l_z$ is the quantum number for the axial component of the angular momentum of the atom.
Since the van der Waals potential is $U_{\mathrm{vdW}}=C_3/r^3$, we have $ |U_{\mathrm{cf}}| \ll |U_{\mathrm{vdW}}|$ when the condition $r\ll r_c\equiv 2mC_3/(\hbar^2|l_z^2-1/4|)$ is satisfied. 
Thus, when the atom is near to the fiber surface and the axial angular momentum is small, the effect of the centrifugal is small compared to that of the van der Waals potential. 
In this case, we can neglect the centrifugal potential. For cesium atoms with $m=132.9$ a.u. and $l_z=10$, the condition for the validity of the above approximation is $r\ll r_c\cong 411$ nm. Next, we note that the reflection of an incident plane wave from a cylindrical surface is rather complicated; it does not produce a plane wave.
However, for the silica--vacuum interface, the reflection coefficient for light with wavelength $\lambda=852$ nm is $R\cong 0.18$, a small number. Therefore, to simplify our calculations, 
we can neglect the reflection of light from the fiber surface. Finally, we note that there are two types of modes of the field in the presence of a nanofiber, namely guided modes and radiation modes. The guided modes are characterized by evanescent waves on the outside of the fiber. They are then the direct analog of the evanescent modes of a dielectric--vacuum interface. 

\begin{figure}[tbh]
\begin{center}
  \includegraphics{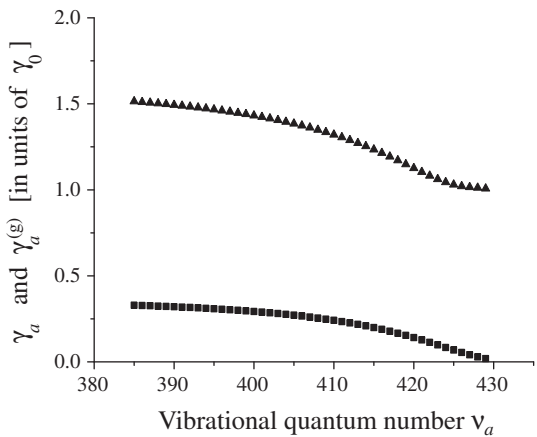}
 \end{center}
\caption{Rate of total spontaneous emission, $\gamma_{a}$ (triangles), and 
rate of spontaneous emission into the guided modes, $\gamma_{a}^{(\mathrm{g})}$ (quadrates),
for a nanofiber. The parameters for the surface--atom potentials are as in Fig.~\ref{fig1}. 
The atomic transition wavelength is $\lambda_0=852$ nm.
The fiber radius is 200 nm. All the rates are averaged with respect to the orientation of the dipole moment. 
}
\label{fig4}
\end{figure}

For a cesium atom at rest, the rate $\gamma(r)$ of total spontaneous emission and the rate $\gamma^{(\mathrm{g})}(r)$ of spontaneous emission into the guided modes of a nanofiber have been calculated systematically in Ref. \cite{cesium decay}.
For an atom in a translational state $|a\rangle$, the rate $\gamma_a$ of total spontaneous emission 
and the rate $\gamma_a^{(\mathrm{g})}$ of spontaneous emission into the guided modes 
can be estimated as \cite{boundspon}
\begin{eqnarray}
\gamma_{a}&=&\int_{0}^\infty \gamma(r)|\varphi_a(r)|^2dr,
\nonumber\\
\gamma_{a}^{(\mathrm{g})}&=&\int_{0}^\infty \gamma^{(\mathrm{g})}(r)|\varphi_a(r)|^2dr.
\label{e8}
\end{eqnarray}
We plot in Fig. \ref{fig4} the rates  $\gamma_{a}$  and  $\gamma_{a}^{(\mathrm{g})}$ 
for a nanofiber with radius of 200 nm, which was used in the experiment \cite{Kali}.
Figure \ref{fig4} shows that the emission into the guided modes is substantial for a wide range of bound 
excited-state levels
($\nu_a<430$). However, when $\nu_a$ is very large ($\nu_a\geq430$), 
the channeling of emission into the guided modes
becomes negligible. The reason is that, when $\nu_a$ is very large 
the bound state is very shallow. The atom in such state spends most of its time far away from the surface, where the evanescent waves in the guided modes  cannot penetrate into.

%%%%%%%%%%%%%%%%%%%%%%% Figure 4

\begin{figure}[tbh]
\begin{center}
  \includegraphics{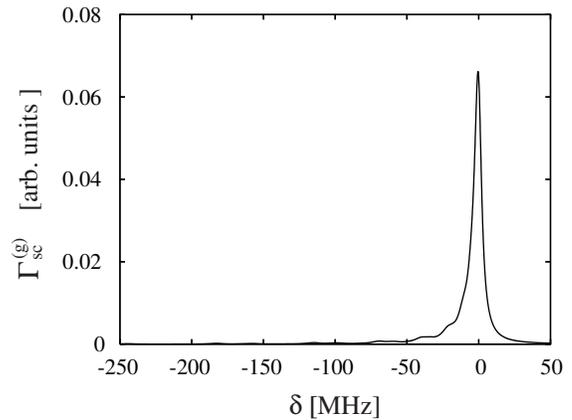}
 \end{center}
\caption{Rate $\Gamma_{\mathrm{sc}}^{(\mathrm{g})}$ of light scattering from the atom into the guided modes of a nanofiber
through the free-to-bound transitions  
as a function of the detuning $\delta=\omega-\omega_0$ of the probe field. The atom is initially in a thermal 
mixture of free ground states with temperature  $T=200$ $\mu$K.
The range of the quantum numbers $\nu_a$ of the bound excited states involved into the calculations is from 385 to 429. The natural linewidth of the atom is
$\gamma_0=5.25$ MHz. Other parameters are as in Figs. \ref{fig1} and \ref{fig4}. 
}
\label{fig5}
\end{figure}

\begin{figure}[tbh]
\begin{center}
  \includegraphics{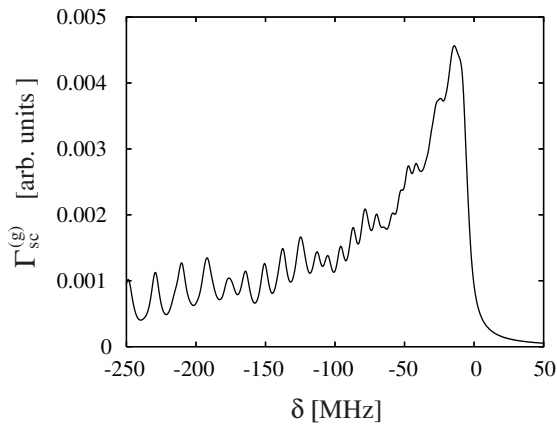}
 \end{center}
\caption{Rate $\Gamma_{\mathrm{sc}}^{(\mathrm{g})}$ of light scattering from the atom into the guided modes of a nanofiber through the bound-to-bound transitions  
as a function of the detuning $\delta=\omega-\omega_0$ of the probe field. 
The atom is initially in an incoherent mixture of a number of bound ground-state levels with quantum numbers $\nu_b$ from 269 to 293 and equal weight factors. 
The bound excited-state levels involved into the calculations are those with the quantum numbers $\nu_a$ from 385 to 429. The natural linewidth of the atom is
$\gamma_0=5.25$ MHz. Other parameters are as in Figs. \ref{fig1} and \ref{fig4}. }
\label{fig6}
\end{figure}

With the above simplifications and approximations, the rate of scattering
into the guided modes of a nanofiber can be estimated by 
\begin{equation}
\Gamma_{\mathrm{sc}}^{(\mathrm{g})}=\frac{s}{2}
\sum_{ab}\frac{\gamma_0^2}{\gamma_a^2+4\delta_{ab}^2}\gamma_a^{(\mathrm{g})}
|F_{ab}(k)|^2\rho_{bb}.
\label{e9}
\end{equation}
The dependence of $\Gamma_{\mathrm{sc}}^{(\mathrm{g})}$ on the field--atom detuning 
$\delta=\omega-\omega_0$ (the difference between the probe field frequency $\omega$ and the free-space atomic resonance frequency $\omega_0$) characterizes the optical excitation spectrum of the atom in the surface-induced potential. We calculate $\Gamma_{\mathrm{sc}}^{(\mathrm{g})}$ for
two cases. In the first case,  the atom is initially in free ground-state levels. In the second case, the atom is initially in bound ground-state levels. 
We plot the results for the first and second cases in Figs. \ref{fig5} and \ref{fig6}, respectively. 

%%%%%%%%%%%%%%%%%%%%%%% Figure 5

In the first case, i.e. the case of Fig. \ref{fig5}, we assume that the initial state
of the atom is  a thermal mixture of free ground states $|b_f\rangle$, 
described by the density operator 
\begin{equation}
\rho(0)=|g\rangle\langle g|\int_0^{\infty}P_E|E\rangle\langle E|\,dE.
\label{e10}
\end{equation}
Here, $|E\rangle$ are the continuum eigenstates, normalized to the delta function of energy, for
the ground-state Hamiltonian $p^2/2m+V_g(r)$,
and $P_E=e^{-E/k_BT}/Z$
is the Boltzmann weight factor, with $T$ and $Z$ being the temperature 
and partition function, respectively, for the atomic system.
The dependence of the scattering rate $\Gamma_{\mathrm{sc}}^{(\mathrm{g})}$ on the field detuning $\delta$ in Fig. \ref{fig5} represents the optical excitation spectrum for the atomic free-to-bound transitions. This spectrum
shows a small negative shift of about $-0.4$ MHz for the position of the peak.
The linewidth of the spectrum is about 6.7 MHz, slightly larger than the natural linewidth 
$\gamma_0=5.25$ MHz of the $D_2$ line of atomic cesium.
In addition, the spectrum has a small, short tail on the negative side of the detuning.
These features are the results of the negative frequency shifts of the transitions between  free ground-state levels and bound excited-state levels.  The observed effects are however weak, and the excitation spectrum is basically concentrated around the atomic resonance
frequency $\omega_0$. The reason is the following:
The Frank--Condon factor for the overlap between a free ground-state and a deep bound excited-state level is small (see Fig. \ref{fig3}). Therefore, a free ground-state atom can be excited only to shallow bound levels (and free levels) of the excited state. Meanwhile, an atom
in a shallow (or free) excited level cannot emit photons efficiently into the guided modes
because the atom spends most of its time far away from the fiber (see Fig. \ref{fig4}).
Hence, the effect of the surface-induced potential on the excitation spectrum
in the case of free-to-bound transitions is rather weak.

%%%%%%%%%%%%%%%%%%%%%%% Figure 6 

In the second case, i.e. the case of Fig. \ref{fig6}, we assume that the initial state
of the atom is in an incoherent mixture of a number of bound ground-state levels $\nu_b$, with
flat weight factors $\rho_{bb}=\mathrm{const}$. The initial density operator of the atom is given by 
\begin{equation}
\rho(0)=\frac{1}{N}\sum_{\nu_b=\nu_{\mathrm{min}}}^{\nu_{\mathrm{max}}} |b\rangle\langle b|,
\label{e12}
\end{equation}
where $N$ is the number of bound ground-state levels involved in the initial state of the atom. The use of flat weight factors is explained by the assumption that the atom is adsorbed by a surface at the room temperature $T= 300$ K $\cong 6.25$ THz, which is much higher than the energy differences between the levels. 
For our numerical calculations, we include the bound ground-state levels with quantum numbers $\nu_b$ from 269 to 293. The energy shifts of these levels, equal to their total center-of-mass energies, are in the range from $-1$ GHz to $-5$ MHz. We do not include levels
with larger quantum numbers because they are too shallow and therefore, an atom in such a bound ground-state level can be easily excited to a free ground-state level by heating or collision. We also do not include levels with smaller quantum numbers  because the corresponding transition frequency shifts are beyond the range of the excitation spectrum measured in the 
experiment \cite{Kali}. 
The dependence of the scattering rate $\Gamma_{\mathrm{sc}}^{(\mathrm{g})}$ on the field detuning $\delta$ in Fig. \ref{fig6} represents the optical excitation spectrum for the atomic bound-to-bound transitions. 
This spectrum
shows a substantial negative shift of about $-14.3$ MHz for the position of the peak.
The linewidth of the spectrum is about 58.3 MHz, which is one order larger than the natural linewidth $\gamma_0=5.25$ MHz of the $D_2$ line of atomic cesium.
Furthermore, the spectrum has a substantial long tail on the negative side
and a small short tail  on the positive side  of the detuning.
These features are the results of the frequency shifts of the transitions between 
bound ground-state levels and bound excited-state levels.
In general, both negative and positive shifts can come into play. 
However, the transitions with positive shifts are relatively weak because the corresponding Frank--Condon factors are usually small. The overlap between the center-of-mass wave functions of a bound excited-state level
and a bound ground-state level is substantial only when the corresponding returning points
are close to each other. Since the van der Waals potential for the excited state $|e\rangle$
is deeper than that for the ground state $|g\rangle$, the frequency shifts of the transitions
that are associated with substantial Frank--Condon factors are mostly negative. 
The deeper the bound ground-state levels involved in the initial state, the stronger the  
effect of the surface--atom interaction on the excitation spectrum.
We note that the basic features of the spectrum in Fig. \ref{fig6} are very
similar to the features of the right and left wings  of 
the experimental spectrum reported in Ref. \cite{Kali}.
However, it is important to recall that the experimental spectrum \cite{Kali} 
has contributions from both bound-to-bound and free-to-bound types of transitions. 
A combination of both Figs. \ref{fig5} and \ref{fig6} with proper weight factors can indeed lead to an excellent matching of the experimental spectrum with the theoretically predicted one.

In conclusion, we have studied the optical excitation spectrum of an atom in the vicinity of a dielectric surface. We have derived rigorous expressions for the rates of the
total scattering and the scattering into the evanescent modes. 
With a proper assessment of the limitations, our theoretical results have been applied to an experimental situation  with a nanofiber.
We have shown that the effect of the surface-induced potential on the excitation spectrum
in the case of free-to-bound transitions is rather weak. 
Meanwhile, the spectrum of excitation of bound-to-bound transitions
has a large linewidth, a substantial negative shift of the peak position,
and a substantial long tail on the negative side
and a small short tail  on the positive side of the field--atom frequency detuning.
The basic features of the calculated spectra are in agreement with the experimental observations \cite{Kali}.

\begin{acknowledgments}
We thank M. Ori\'{a}, K. Nayak, and P. N. Melentiev for fruitful discussions.
This work was carried out under the 21st Century COE program on ``Coherent Optical Science.''
\end{acknowledgments}

\end{document}